\documentclass{article}

\usepackage{arxiv}

\usepackage[utf8]{inputenc} % allow utf-8 input
\usepackage[T1]{fontenc}    % use 8-bit T1 fonts
\usepackage{hyperref}       % hyperlinks
\usepackage{url}            % simple URL typesetting
\usepackage{booktabs}       % professional-quality tables
\usepackage{amsfonts}       % blackboard math symbols
\usepackage{nicefrac}       % compact symbols for 1/2, etc.
\usepackage{microtype}      % microtypography
\usepackage{lipsum}		    % Can be removed after putting your text content
\usepackage{graphicx}
\usepackage[numbers,square]{natbib}
\usepackage{doi}
\usepackage{amsmath}
\usepackage[bottom]{footmisc}
\usepackage{footnote}
\hypersetup{hidelinks}
\usepackage{lipsum}

\title{Facial Expression Re-targeting \\from a Single Character}

\date{}

\makeatletter
\def\@fnsymbol#1{\ensuremath{\ifcase#1\or \dagger\or *\or \ddagger\or
   \mathsection\or \mathparagraph\or \|\or **\or \dagger\dagger
   \or \ddagger\ddagger \else\@ctrerr\fi}}
\makeatother

\author{\href{https://orcid.org/0000-0002-5006-9300}{\includegraphics[scale=0.06]{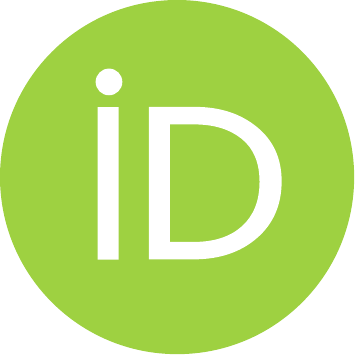}\hspace{1mm}Ariel Larey\thanks{Authors contributed equally to this work.}} \\
    Huawei Research, Israel\\
		\And
  \href{https://orcid.org/0000-0003-4644-6545}{\includegraphics[scale=0.06]{orcid.pdf}\hspace{1mm}Omri Asraf$^{\dagger}$} \\
	Huawei Research, Israel\\
		\And
  \href{https://orcid.org/0000-0002-4756-3922}{\includegraphics[scale=0.06]{orcid.pdf}\hspace{1mm}Adam Kelder}\\
    Huawei Research, Israel\\
	    \And
     \href{https://orcid.org/0000-0003-0915-7760}{\includegraphics[scale=0.06]{orcid.pdf}\hspace{1mm}Itzik Wilf} \\
	Huawei Research, Israel\\
		\And
  \href{https://orcid.org/0000-0002-6963-3645}{\includegraphics[scale=0.06]{orcid.pdf}\hspace{1mm}Ofer Kruzel} \\
	Huawei Research, Israel\\
		\And
	\href{https://orcid.org/0000-0002-0939-3379}{\includegraphics[scale=0.06]{orcid.pdf}\hspace{1mm}Nati Daniel$^{\dagger,}$\thanks{Corresponding author, e-mail: snatidaniel@gmail.com.}} \\
	Huawei Research, Israel
}

\renewcommand{\headeright}{Larey A et al.}
\renewcommand{\shorttitle}{Facial Expression Re-targeting from a Single Character}

\begin{document}

\maketitle

\begin{abstract}
Video retargeting for digital face animation is used in virtual reality, social media, gaming, movies, and video conference, aiming to animate avatars' facial expressions based on videos of human faces. The standard method to represent facial expressions for 3D characters is by blendshapes, a vector of weights representing the avatar's neutral shape and its variations under facial expressions, e.g., smile, puff, blinking. Datasets of paired frames with blendshape vectors are rare, and labeling can be laborious, time-consuming, and subjective. In this work, we developed an approach that handles the lack of appropriate datasets. Instead, we used a synthetic dataset of only one character. To generalize various characters, we re-represented each frame to face landmarks. We developed a unique deep-learning architecture that groups landmarks for each facial organ and connects them to relevant blendshape weights. Additionally, we incorporated complementary methods for facial expressions that landmarks did not represent well and gave special attention to eye expressions. We have demonstrated the superiority of our approach to previous research in qualitative and quantitative metrics. Our approach achieved a higher MOS of 68\% and a lower MSE of 44.2\% when tested on videos with various users and expressions.
\end{abstract}


\section{Introduction}
Various applications use video retargeting for digital face animation. These domains include social media, gaming, movies, and video conferences. Video retargeting systems aim to translate human facial expressions into 3D characters, eventually mimicking the real footage expressions. 

A common method to represent facial expressions of 3D characters is by blendshapes. In this method, different mesh shapes serve as targets where each mesh target has its corresponding blendshape weight that determines its significance in the desired expression. Mathematically, the mesh targets serve as eigenvectors and the blendshape weights as linear combination coefficients, resulting in an interpolated expressed mesh.
In the case of video retargeting, the objective is to translate expressions from real footage videos to sequences of blendshape weights to control 3D characters' facial animation. Particularly in this study, we used a 3D character with 62 mesh targets with semantic meanings such as smile, puff, and blink in addition to multiple Visemes. 

The large variation of facial expressions and character shapes poses a significant obstacle in training efficient models: creating a large, representative dataset of video frame - blendshape weights pairs. This dataset type can be created manually by labeling each video frame with its blendshape weights. 

Yet, such a process is labor-intensive, time-consuming, and highly subjective – when individual observers can interpret the same image differently, causing a nondeterministic labeling process. A reasonable solution is training the machine learning model using a synthetic dataset. 

\begin{figure}[ht!]
  \centering
  \includegraphics[width=\linewidth]{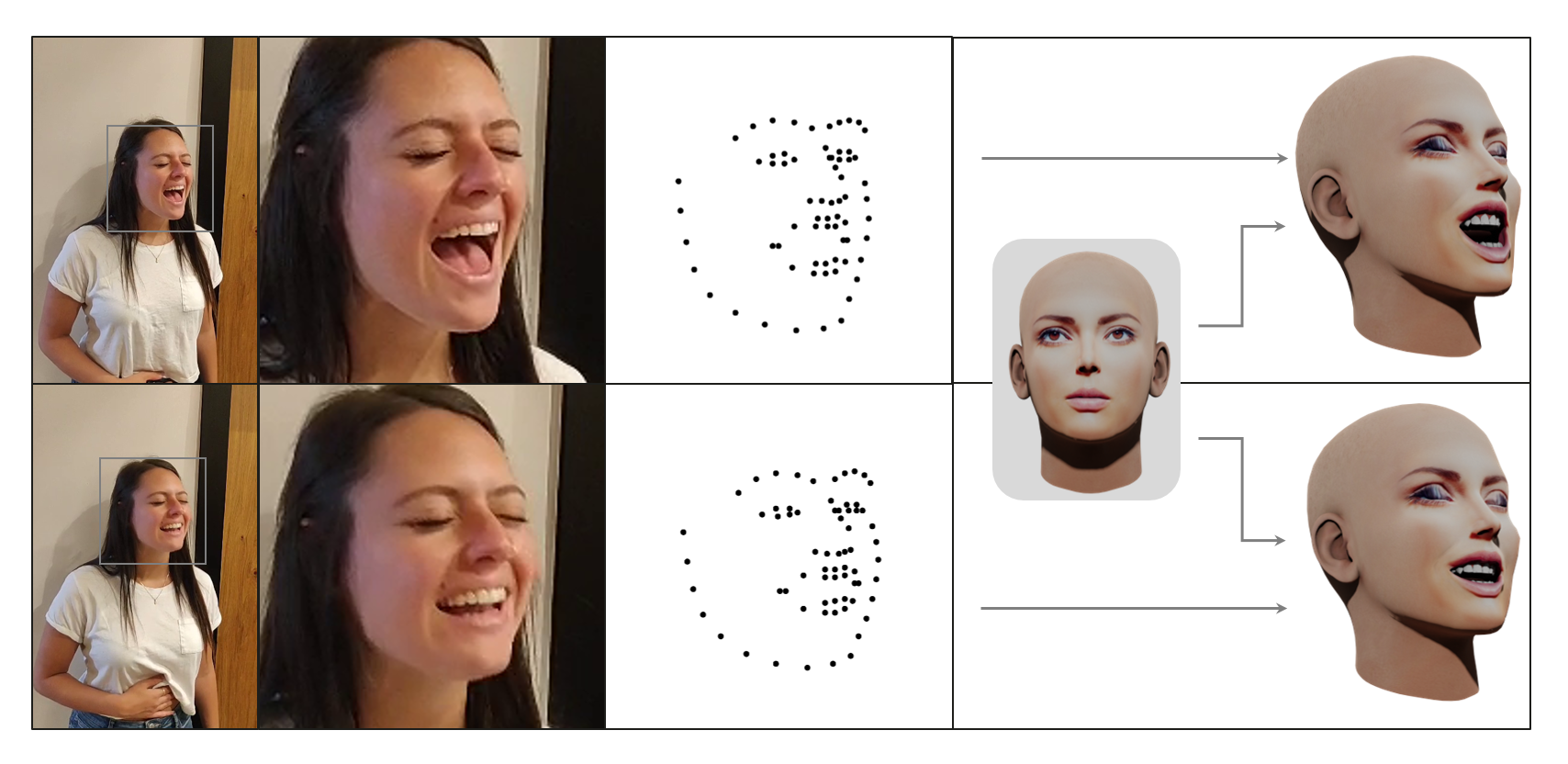}
  \caption{Results of our method on two video frames. From left-to-right: input image, cropped input image, extracted facial landmarks, our output expressed 3D character.
}
\label{fig0}
\end{figure}

In this approach, realistic 3D characters deform into numerous expressions based on pre-defined blendshape weights, which serve as the dataset's deterministic ground truth. Next, each mesh is rendered into realistic scenes used as input images during the training procedure. However, a significant challenge in this approach is overcoming the domain gap between the synthetic scenes the model encounters during training and the real scenes it encounters during inference. Producing a diverse synthetic dataset using high-poly realistic 3D characters that represent the photorealistic scenes sufficiently can, on the one hand, narrow the domain gap but, on the other hand, can be very expensive.   
We combat that multiplicative growth of efforts by training the video to blendshape weights conversion on a single character. To narrow the domain gap between the trained single character to the realistic human facial scenes, we use a well-known representation of face structure that captures expressions quite well – facial landmarks, particularly the standard set of 68 landmarks \citep{kazemi2014one}. The conversion of a face image, depicting an actual person or a human-like ("realistic") 3D character, to a set of landmarks performs data reduction into a "symbolic" representation. We train a blendshape translation network in that space using a single character.

Furthermore, the local nature of both blendshape coefficients and face landmarks allows partitioning the problem into subsets – e.g., eye landmarks versus eye blendshapes. Moreover, working with blendshape provides an additional advantage – the ability to apply the predicted blendshape coefficients to characters of identical topology but different geometry, thus generalizing from our single character to a wide range of avatars.

Section 2 reviews related work, highlighting specific challenges. 
Our method is described in section 3, starting with our data preparation. We generate synthetic data according to a real-world distribution of head poses and by modifying blendshape coefficients to generate plausible expressions from a single 3D character that can be matched to video frames of real actors.

We further apply landmark detection to the synthetic images and describe how to regress blendshape weights locally from landmarks. While the landmark networks reliably present many expressions, they fail to replicate eye blinks and eye gaze directions. Other expressions, like puff and sneer, occur in face regions, poorly represented by landmarks. Thus, we added complementary methods for these regions and expressions.

Finally, our results in section 4 show that our approach scales well to videos with different users and various expressions, outperforming previous research in qualitative and quantitative metrics (Figure~\ref{fig0}).

\section{Related Work}
Video-driven facial animation, particularly in the context of Video retargeting / Motion capture, is a challenging task that has gained significant attention in recent years. One of the primary objectives of video-driven facial animation is to automatically transfer facial expressions from videos onto 3D characters. 

Hence, it enables a wide variety of product applications in virtual reality and augmented reality eras. In particular, in social media, gaming, movies, music clips, and video conference scenarios, driving a 3D avatar gives the illusion of the real world and improves the viewer and user experience. 

Several techniques and approaches have been proposed to realistically address the complexities of capturing and transferring facial expressions. In this section, we present a review of the relevant literature and highlight the key contributions and advancements in the field.

\subsection{Facial Expressions Animation Transfer}
The emergence of deep learning has revolutionized many areas of computer vision and computer graphics, including facial animation transferring. Many researchers have leveraged deep neural networks, generative adversarial networks, and 2D/3D facial landmarks for facial animation retargeting, whose goal is typically to capture the facial performances of the source actor and then transfer the expression to a target character.

\citep{cao20133d} suggests a system that learns to regress blendshapes values based on 3D facial landmarks. This approach used a paired database of 2D images and user-specific blendshapes for the training process. \citep{Siarohin_2019_CVPR} developed a novel deep model based on a generative adversarial network (GAN) \citep{goodfellow2020generative} that transfers the captured source facial expressions to a different actor, thus allowing for personalized retargeting.

Furthermore, \citep{cao2014displaced} suggests a deep learning solution that regresses a displacement map to predict dynamic expression based on inferring accurate 2D facial landmarks and the geometry displacements from an actor video. \citep{cao2018stabilized} derived a system that learns the dynamic rigidity of prior images from 2D facial landmarks and motion-guided correctives \citep{kroeger2016fast}. In addition, \citep{barros2019face} developed a method for 2D-3D facial expression transfer by estimating the rigid head pose and non-rigid face expression from 2D actor facial landmarks using an energy-based optimization solved by the non-linear least square problem. \citep{peng2023facial} introduced a method that combines optical-low estimation with a mesh deformation model to establish correspondences between the given actor video to the target 3D character.

Besides, other approaches have recently been using domain transfer methods such as \citep{moser2021semi, aneja2018learning} to animate any 3D characters from human videos. Specifically, \citep{aneja2018learning} method learns to predict the facial expressions in a geometrically consistent manner and relies on the 3D rig parameters. This requires a large database of synthetic 2D image characters aligned to 3D facial rig \citep{dhall2011static, lucey2010extended, mavadati2013disfa, pantic2005web}. While the \citep{moser2021semi} method learns to transfer animations between distinct 3D characters without consistent rig parameters and any engineered geometric priors.

\subsection{Facial Expressions Animation Synthesis}
Recently, there has been a shift towards 3D-based approaches for facial expression synthesis. With the availability of depth sensors and advanced 3D modeling techniques, researchers have explored methods to capture and represent facial expressions in three dimensions. This has led to the development of more accurate and detailed facial expressions models, such as 3D Morphable Models (3DMMs) and blendshape-based models, which enable more realistic retargeting of facial expressions onto different actors \citep{egger20203d, sanyal2019learning, chaudhuri2019joint, chaudhuri2020personalized}.

\section{Method}
\subsection{Preliminaries}
Our method relies on datasets with reasonable facial expressions and with natural head poses, similar to the way achieved in our approach. The uniqueness of this process is that the main training is performed via only one character that is required to be a polygonal mesh of fixed topology that deforms to different blendshape targets linearly.

Moreover, we assume the character is a realistic avatar with human facial geometry to reduce the domain gap between the training character and the inferred real actors. The blendshape coefficients predicted during inference could also be applied to other characters, even to stylized avatars with unnatural geometries. Yet, we assume these characters support the same blendshape targets as the original character used for training. In addition, we assume these blendshape targets are constructed with a semantic meaning.  

Furthermore, our methodology does not require any temporal information in both domains. On the one hand, we do not use any information regarding the character's animation or continuously rendered frames for training. On the other hand, our pipeline could work during inference on actors’ frames that lack temporal relation.

\begin{figure}[h]
  \centering
  \includegraphics[width=\linewidth]{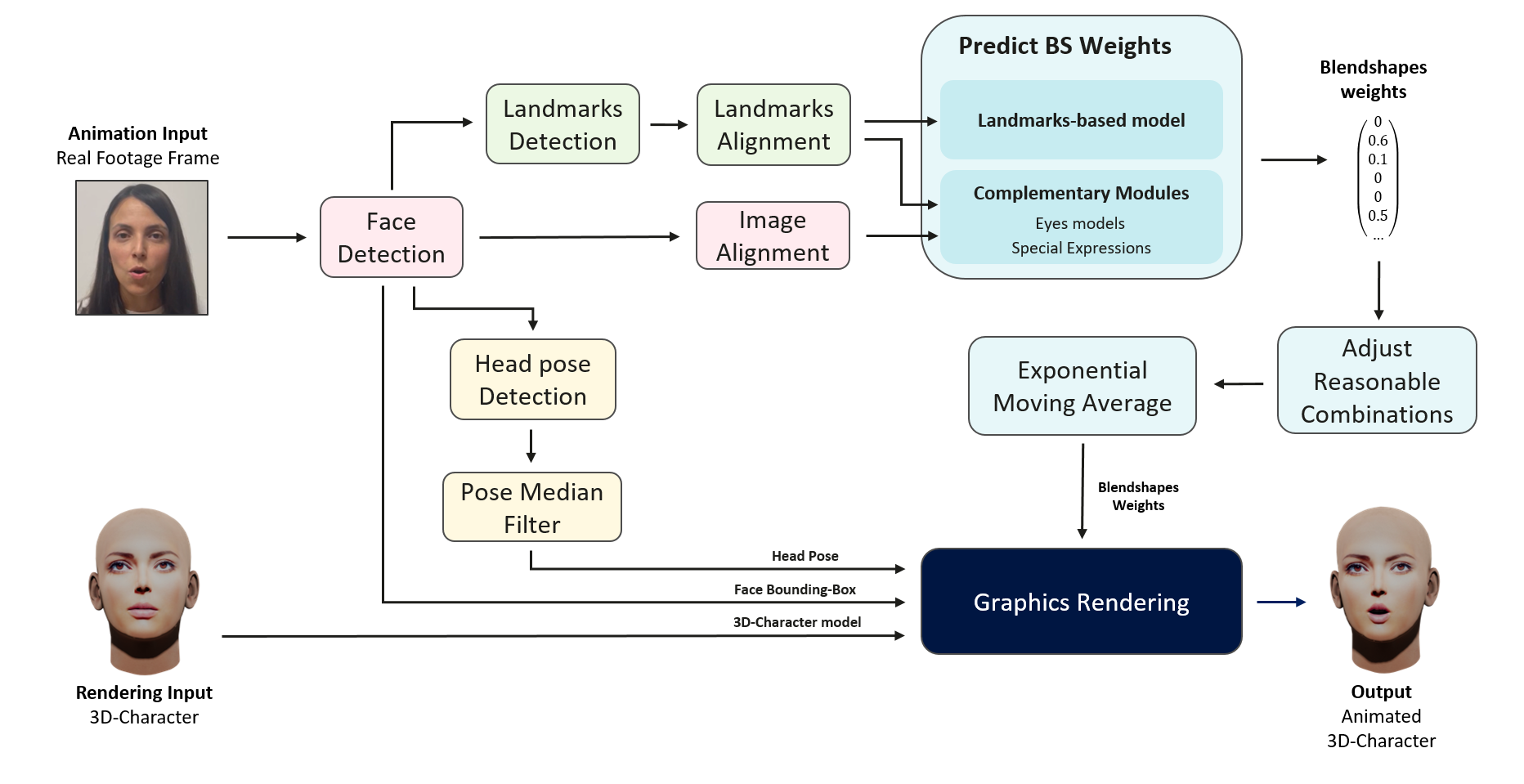}
  \caption{Overview of Facial Expression Re-targeting inference pipeline. The full solution includes sub-modules applied to real footage, such as face detection to the given video frame and extracting its face landmarks as a pre-processing phase. Then, both face images and landmarks go through an alignment process. The given aligned face and landmarks are the input for Deep convolutional Neural networks (Landmarks to blendshapes weights network and complementary networks) that predict the corresponding blendshape weights of the given character model. The predicted weights are post-processed with the corrective expressions method (Reasonable Combinations) and stabilized based on previous frames' information (Exponential moving average). Finally, a Character is then rendered from real footage head pose, and the predicted blendshape weights to demonstrate the animated 3D character.}
\label{fig1}
\end{figure}

\subsection{Pipeline architecture}
Our pipeline comprises several building blocks that process a facial image and predict its corresponding blendshape weights. The main route starts by detecting the face boundaries using a face detection model \citep{bulat2017far} and cropping the image based on the bounding box. Next, facial landmarks are extracted by a pre-trained model as well \citep{bulat2017far}. As a final pre-process step, the landmarks are aligned into a frontal position with a resolution of 128X128 pixels by an sRT transformation.

The aligned landmarks are used as the input to a dedicated deep model that predicts the blendshape weights that serve as the coefficients for the blendshapes linear combination. As a post-process, we fine-tune the predicted weights to verify that they are plausibly visible. To accomplish that, we constructed in advance an array (coined Reasonable-Array) that consists of all blendshape target pairs. The Reasonable Array provides a binary indication of whether each pair of targets should be enabled together. When a pair of targets is not reasonable, the weight prediction of the smaller target is zeroed. Furthermore, When the input frames are part of a video sequence, they are processed by an Exponential Moving Average operation to smooth the temporal dynamics. Moreover, the head pose of the actor is predicted via Hope-net \citep{doosti2020hope} for rendering orientation knowledge. 

Yet, some targets are prone to errors when predicting them using only landmarks. Eyes expressions detection using landmarks is a challenging task where the landmarks' prediction errors propagate to the final predictions. Moreover, for some blendshape targets, such as Puff and Sneer expressions, only 68 standard landmarks fail to cover relevant facial regions. Subsequently, the model miss-predicts the corresponding facial expressions. Thus, To achieve weights prediction over the entire set of blendshapes, complementary modules are being used in the challenging cases where the landmarks model fails. During these special scenarios, the cropped images are aligned into 128X128 pixels resolution and serve as an input to the complementary modules.

Finally, the predicted blendshape weights, face bounding box, and head pose are applied to a 3D-mesh rendering procedure. Fig. \ref{fig1} illustrates the overall pipeline for inferring a 2D image, predicting its blendshape weights, and re-targeting them to a 3D mesh.

\begin{figure}[h]
  \centering
  \includegraphics[width=\linewidth]{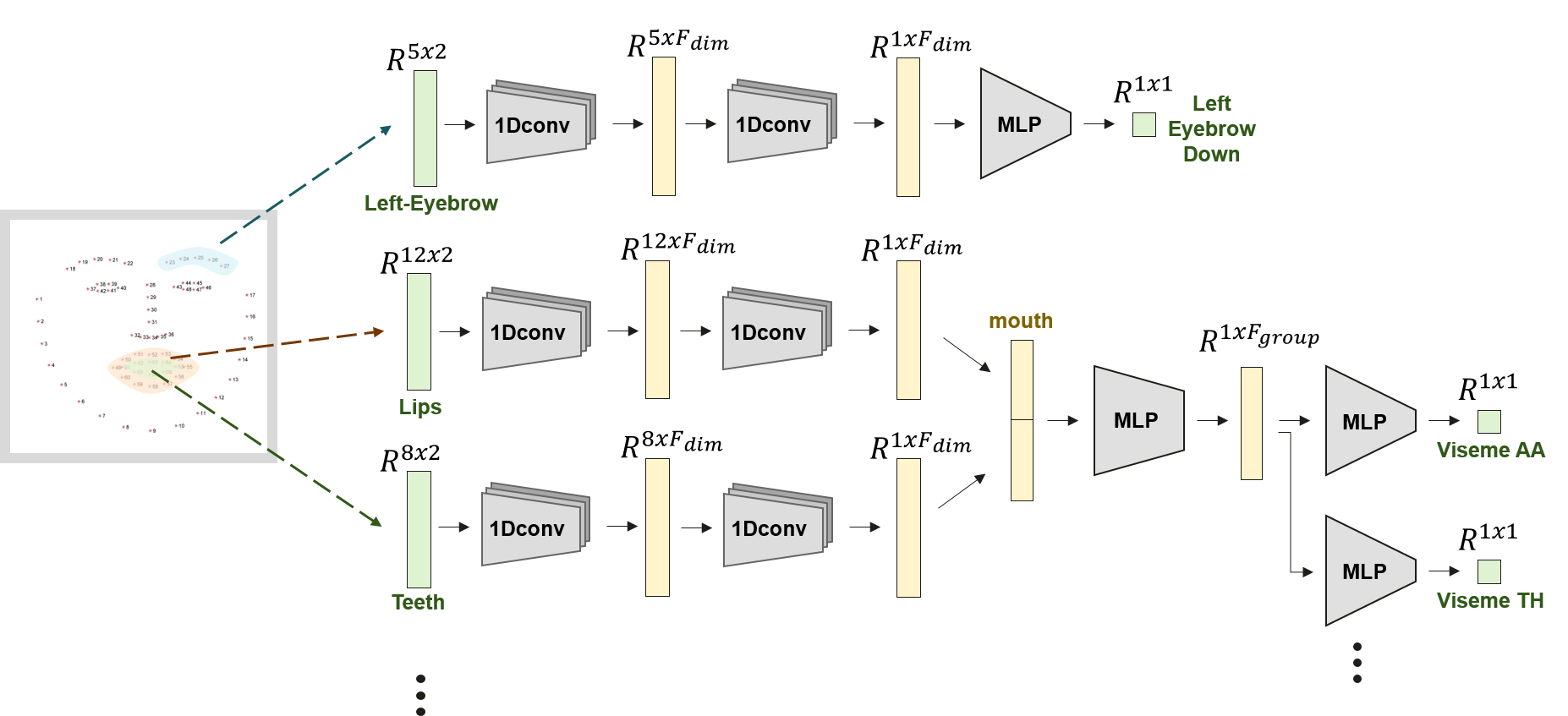}
  \caption{Landmarks to blendshape weights network architecture. The architecture stages: (1) Separating the landmarks into facial regions. (2) Feature extraction for each region by 1D-convolution layers. (3) Grouping and connecting the regions to the relevant handshape weights. }
  \label{fig2}
\end{figure}

\subsection{Data Preparation}
The main deep learning model was trained using a single 3D character. The objective of the data preparation phase is to create 2d landmarks of the character in various head-poses and reasonable expressions while maintaining its blendshape weights as the ground truth for the training procedure. 

\subsubsection{Synthetic Data Pre-process}
First, we learn the distribution of natural head poses by predicting the orientation of real-life video frames that reflect the use-case scenario. The head pose prediction is performed by Hope-net \citep{doosti2020hope}, and the pose of each frame is stored in a collection of realistic head poses.

Furthermore, the procedure requires creating manually in advance a Reasonable-Array that is adjusted to the character that defines blendshapes pairs that could be enabled simultaneously.

\subsubsection{Synthetic Data Creation}
To create pairs of landmarks and ground truth blendshape weights for the training procedure, we start by rendering the 3D character into 2D images in various poses and expressions.

The head pose is randomly sampled from the realistic head poses constructed in the pre-processing phase. The expression of each frame is generated randomly, with the following limitations: (1) No more than five blendshapes are active, where active weights are considered as weights with values larger than zero. (2) Each active blendshape weight ranges from (0,1]. (3) All active weight combinations are reasonable, based on the Reasonable-Array. 

Next, we process the high-resolution rendered images, similar to the process done in the inference pipeline. Still, in this case, it is performed over the rendered character images and not on real actor images. First, we obtain the bounding box of the character's face, extract its facial landmarks, and finally, align the landmarks into a frontal 128X128 pixels resolution. 

\subsubsection{Real Footage Data Creation}
Based on the real-actor images dataset, specific blendshape targets not accurately reflected by the standard landmarks are predicted differently. Two eyes blinking weights are predicted by a dedicated model. In this case, we capture videos from different actors, where the actors must perform the same eye expression during the entire video, while other facial expressions, head pose, and distance from the camera alter. 

This technique enabled a simple and convenient labeling procedure, where all decoded frames from the same video are labeled the same. Each video contains one of the following eye expressions: (1) close both eyes, (2) natural open eyes, (3) wink, (4) close eyes partly. All videos are decoded into frames where each frame is assigned to the label of the entire video. In addition to the blinking, a dedicated model is trained for 'Puff' and 'Sneer' facial expressions using the same data-collecting technique.

\subsection{Landmarks-to-Weights}
This main module aims to translate the knowledge represented by the facial landmarks to the facial expressions, and Visemes are reflected as the blendshape weights. Thus, the inputs to the model are 68 aligned landmarks, each represented by its two horizontal and vertical coordinates (68X2 shape). In contrast, the model outputs 62 blendshape weights that indicate the coefficients of the linear combination of the blendshape targets.

The simple approaches of training regressors between the source landmarks domain to the target blendshape weights domain did not converge well. However, separating the landmarks into facial regions and propagating the information in a hierarchical route showed high performance. We pre-defined eight landmarks regions: eyebrow-left, eyebrow-right, eye-left, eye-right, nose, nostril, teeth, and lips. 1D-convolution layers processed each region of 2-dimensional landmarks into a 1-dimensional hidden layer representing the region's extracted knowledge. 

The next step depends on the behavior of the blendshape targets. A grouping layer is required when blendshape targets are influenced by more than one landmarks region. In this case, the relevant regions' hidden layers are concatenated and processed by an additional MLP. Finally, each blendshape target has its dedicated MLP that outputs a scalar value in the range of [0,1] that represents the blendshape weight's value. 

For example, as demonstrated in Fig. \ref{fig2}, the target representing the 'AA' Viseme depends on the 'Teeth' and 'Lips' regions of landmarks. Thus 'Lips' and 'Teeth' landmarks regions are grouped into the 'Mouth' group in advance. On the other hand, when the blendshape target is affected directly by only one region of landmarks, the region's hidden layer is regressed directly to predict the corresponding blendshape weight. For example, the blendshape target representing the left lowered eyebrow depends only on the left-eyebrow region of landmarks (Fig. \ref{fig2}).

\subsection{Complementary modules}
\subsubsection{Blink Detection}
Predicting blinking targets given facial landmarks is challenging due to the high diversity of eye structure and surrounding textures that cause landmarks extractors' failures. These errors propagate into the landmarks-based model when predicting blendshape weights. 

Thus, we train a dedicated model to directly predict two blinking blendshape weights from the given image using the real-footage dataset. The input images are aligned and cropped around the eyes to the resolution of 16X40 pixels and applied to a ResNet18 model \citep{RESNET_Kaiming}.

Yet, predictions do not account for different eye geometries. I.e., eyes with narrow geometry could be interpreted as partially closed. Therefore, we adjust the prediction range of values to the individual actors’ eye geometry. For each frame, the distance between the lower eyelid to the upper eyelid is calculated for both eyes, which is then classified by online K-means into one of two classes: 'opened eyes' and 'closed eyes'. K-means averaged values of the two classes are updated during the video progress by the distances calculated per frame. The 'opened eyes' class value reflects the actor's natural eyelids distance and is converted by a linear transformation to a threshold between 0 to 1, which serves as the new low edge for the blinking prediction range of values.          

\subsubsection{Gaze Detection}
Herein, we derived a practical approach for accurately determining and monitoring the direction of an actor's gaze. This involves identifying whether the actor's gaze is directed straight ahead (the Primary position) or in one of the secondary positions, namely up, down, right, or left. 

Our method relies on comprehensive facial eye landmarks, including the iris, inner corner keypoints, and outer corner keypoints. Specifically, the calculations are based on the distance between the iris and the inner and outer corner keypoints.

To predict the coefficients for detecting the direction of the eyes' gaze, we outline the following three steps:
\begin{enumerate}
\item Horizontal Eye Line Calculation: We begin by calculating the properties of the horizontal eye line using the key points of the eye corners. These properties include the mid-point, line slope, bias, and the L2 distance of the horizontal eye.
\item Intersection Point Determination: Next, we determine the intersection point between the iris projection and horizontal eye lines.
\item Secondary Positions Detection: By analyzing the obtained intersection point, we identify the secondary positions of the gaze:
\begin{enumerate}
\item Left and Right Gaze: We measure the distance of the intersection point relative to the mid-point. This distance is normalized based on the individual's horizontal eye L2 distance, resulting in a unique value for each actor. The direction of the eye is correlated with the position of the intersection point relative to the mid-point.
\item Up and Down Gaze: We measure the distance between the intersection and iris points to detect the upward or downward gaze. The eye's direction is correlated with the position of the iris point relative to the horizontal eye line.
\end{enumerate}  
\end{enumerate} 

In summary, leveraging comprehensive facial eye landmarks and employing specific calculations can predict the gaze blendshape coefficients for various eye positions while ensuring smooth and reliable results.

\subsubsection{Special Expressions Detection}
'Puff' and 'Sneer' expressions are part of the Facial Action Coding System (FACS) and refer to facial muscle movements. The standard landmarks are not sufficient to capture these expressions. Thus, to detect the corresponding blendshape weights, we train a ResNet18 model \citep{RESNET_Kaiming}, using the real-footage dataset.

\section{Experiments}
We used three datasets encompassing various facial expressions and video sequences to conduct our experiments. In addition, we demonstrated our study's methodological advancement through reproducibility, transparency, and comparability. Finally, we evaluated our approach's performance using qualitative and quantitative measures. Assessing the quality and performance of facial expressions retargeting algorithm is a crucial aspect of research in this domain. Therefore, various evaluation metrics have been proposed to measure the realism, accuracy, and perceptual quality of retargeted facial expressions, poses, and identities.

\subsection{Datasets}  
\subsubsection{Synthetic Character Landmarks Dataset}
We used a 3D character consisting of 12.8K vertices for each of its 62 blendshape targets to train the landmarks-based model. The blendshapes included various facial expressions and mimics in addition to 14 Visemes (Attribution 4.0 International license \citep{Skullvez}). We created manually in advance a Reasonable Array that is adjusted to this specific set of blendshapes and performed the data preparation as described above. 

We used the Blender tool to render 30,000 images of the character, saved their corresponding blendshapes weights as ground truth, and extracted their aligned facial landmarks. The head pose of each frame scene was sampled from a natural head pose distribution. This distribution was obtained by detecting head pose information from relevant Youtube and Denver Intensity of Spontaneous Facial Action (DIFSA) videos \citep{mohammad2013DIFSA}.

\subsubsection{Real Footage Labeled Dataset}
We collected 200 videos from 40 actors targeting blinking, Puff, and Sneer expressions. The videos were decoded at 30 fps yielding 17101 real footage frames and their corresponding labeled blendshape weights. This dataset was dedicated to supervised training of the Special Expressions models.

\subsubsection{Real Footage Unlabeled Dataset}
To evaluate the performance of our pipeline, we captured additional real-footage videos from 20 identities. The actors were requested to perform various FACS expressions and Visemes. The videos were decoded at 30 fps resulting in 25075 real footage frames.

\subsection{Training Procedure}
The updated model was trained and optimized using Pytorch \citep{Pytorch_framework} framework on a single NVIDIA GeForce GTX 1080 with 24GB GPU memory. 

The hyperparameters of the model that optimize convergence were examined using Adam Solver \citep{Adam_paper} with beta1=0.5 and beta2=0.999, eps of 1e-10, weight decay of 1e-7, a minibatch of size 16, a learning rate of 5e-5. In contrast, we decay the learning rate to zero by gamma=0.5 every three epochs. Weights were initialized from a uniform distribution described in \citep{glorot2010understanding}.
We use Leaky ReLUs with slope 1e-2 for the convolutional layers and a fully connected layer, in addition to Sigmoid activation at the end of the fully connected layer.

The optimization loss function contains two terms. First, a mean square error (MSE) between all ground truth and predicted blendshape weights. Second, we used an MSE between active ground truth and predicted blendshape weights. At the same time, the loss function elements were weighted with values of 1 and 0.1, respectively.

\begin{figure}[h]
  \centering
  \includegraphics[scale=1.1]{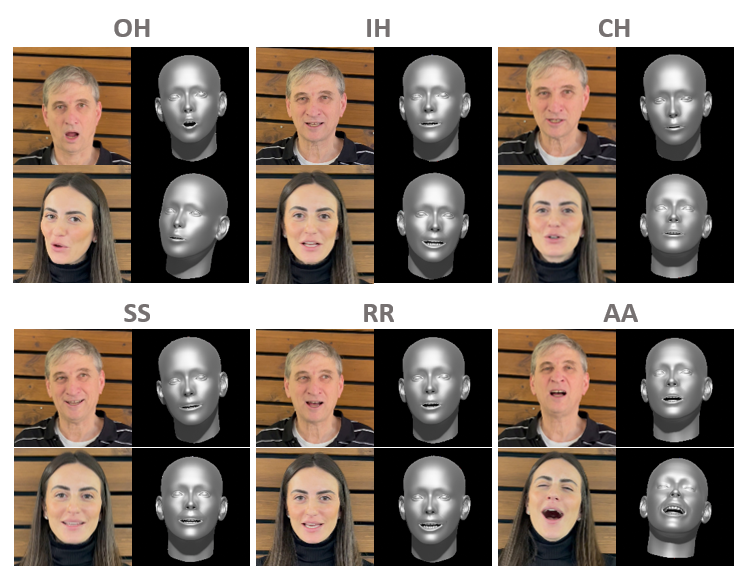}
  \caption{Examples of prediction performance to a variety of Visemes. For each Viseme from left to right: (1) input image, (2) our output expressed 3D character.  }
  \label{viseme}
\end{figure}

\subsection{Results}
\subsubsection{Facial Expressions and Visemes}
\begin{figure}[h]
  \centering
  \includegraphics[width=\linewidth]{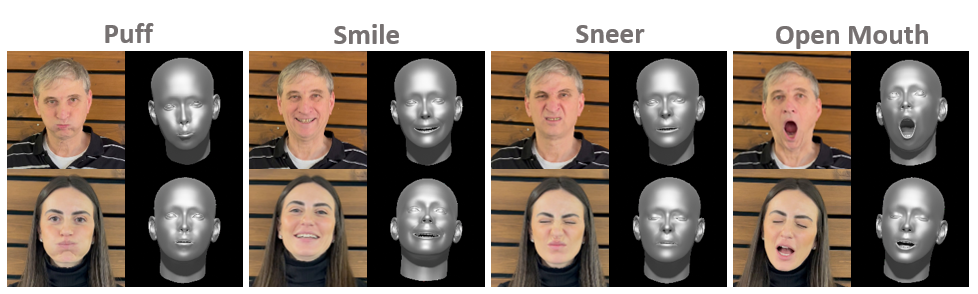}
  \caption{Examples of prediction performance to a variety of expressions. For each expression from left to right: (1) input image, (2) our output expressed 3D character.}
  \label{expresion}
\end{figure}

We examined our method using the Real Footage Unlabeled Dataset. The video frames were subjected to our pipeline, producing Blendshape weights, which serve as coefficients of the linear combination between the mesh geometry targets. Fig. \ref{viseme} and Fig. \ref{expresion} show examples of real footage frames and their corresponding rendered mesh with the translated facial expression. The former presents different Visemes as part of a video speaking sequence, while the latter presents other common facial expressions frames. 

\subsubsection{Multiple Characters}
\begin{figure}[h]
  \centering
  \includegraphics[width=\linewidth]{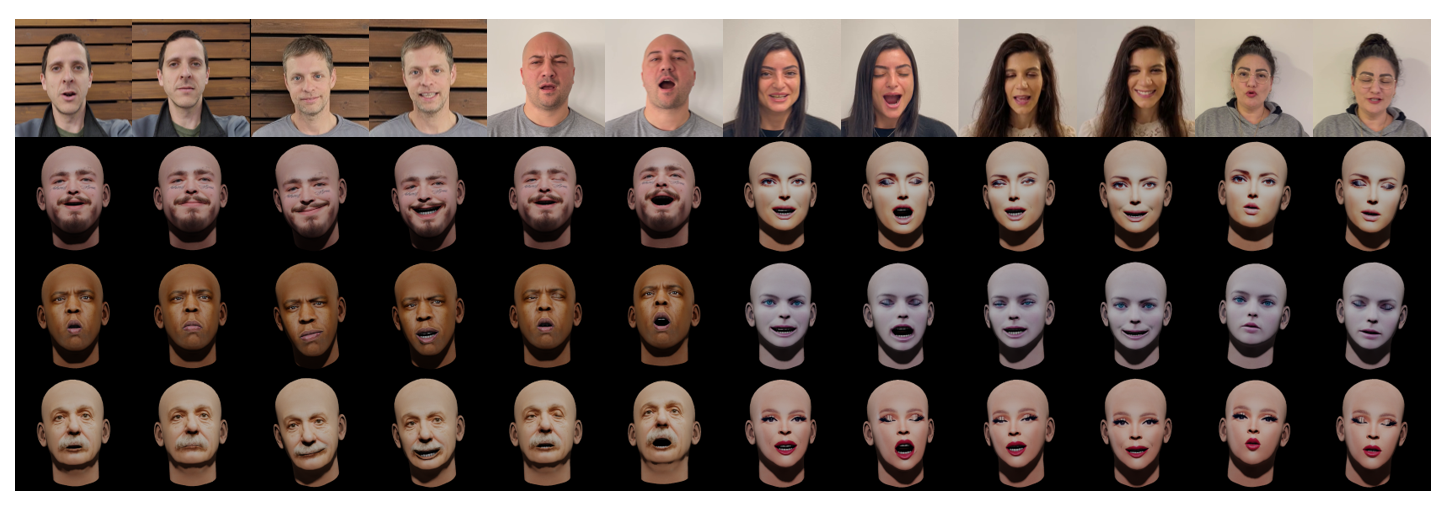}
  \caption{Examples of performance when retargeting to multiple 3D characters from different actors. For each actor from top-to-bottom: (1) input image, (2) our results expressed on three different characters.}
  \label{multy}
\end{figure}

The uniqueness of our method stems from the single-character training procedure where only one 3D character is enough to obtain sufficient results. Yet, we can drive any desired character during inference if it supports the same geometrical space and can be animated using the same semantic blendshapes. Fig. \ref{multy} shows the robustness of our model over multiple actor identities and multiple 3D characters constructed by their unique geometries and textures. These characters were not part of the training but are represented in the same semantic blendshape space as the single mesh used for training \citep{Skullvez}. 

\subsubsection{Competitive comparison}
\begin{figure}[h]
  \centering
  \includegraphics[scale=0.5]{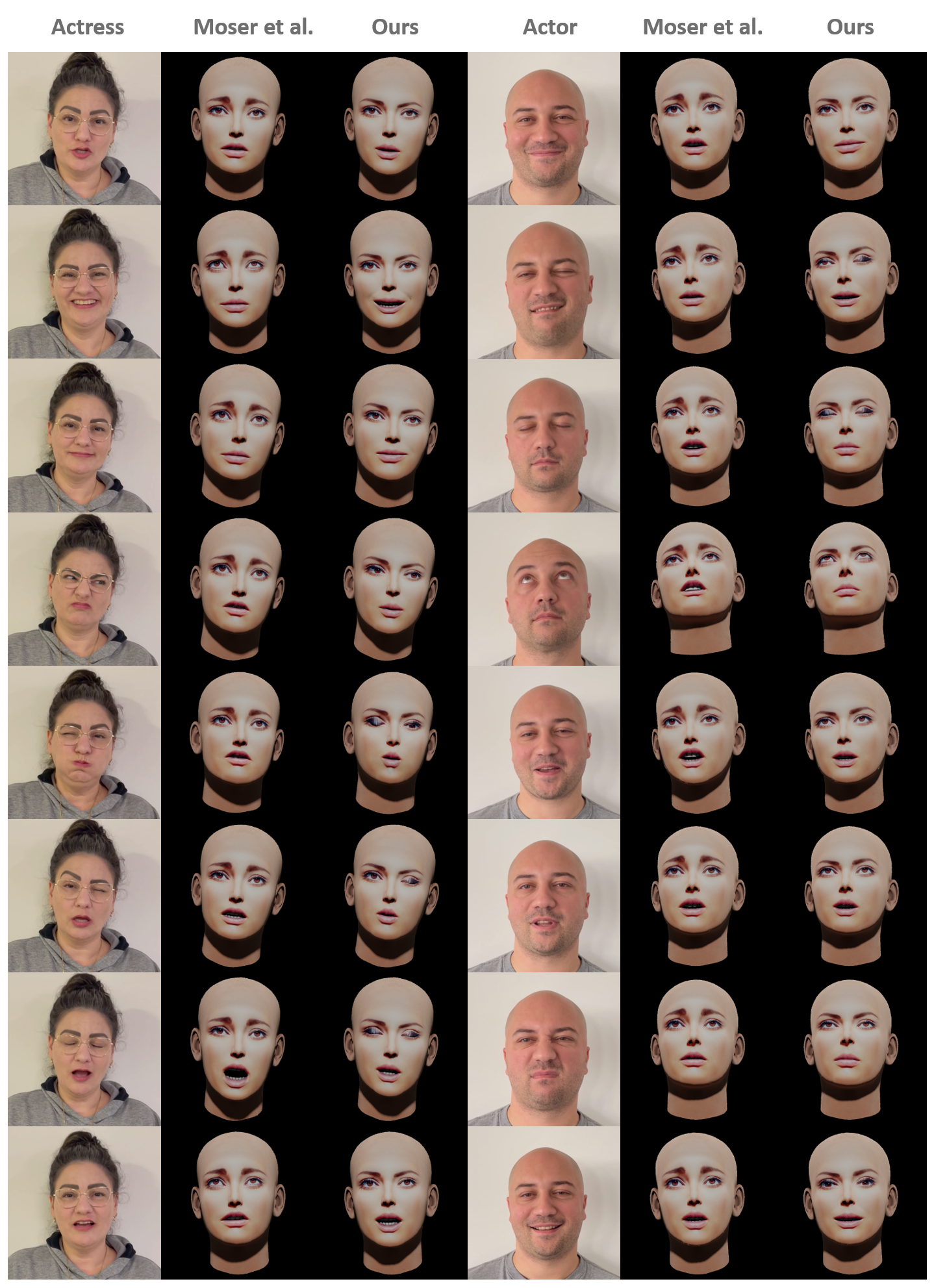}
  \caption{Video retargeting comparing. Examples of ours vs. Moser et al. pipeline results of different facial expressions for two real actors.}
  \label{comparison}
\end{figure}

As a baseline to the single-character video retargeting pipeline, we implemented the algorithm proposed in \citep{moser2021semi} where a semi-supervised approach that included an image translation technique was introduced. We inferred the baseline model over the Real Footage Unlabeled Dataset as well. Examples of our method's results compared to \citep{moser2021semi} method are presented in Fig. \ref{comparison}.

For qualitative evaluation, we conducted a subjective user study where 80 individuals were required to rate the degree of compatibility between pairs of actors and their corresponding rendered 3D character images in facial expressions aspects. The user study contained 34 pairs, where each actor frame appeared twice, once with a prediction obtained by our method and once with a prediction obtained by \citep{moser2021semi} method (each time in random order). The subjects rated each pair on the Likert scale (scores between one to five), and we reported their Mean Opinion Score (MOS) separately for Visemes representative frames and other facial expressions representative frames.

\begin{table*}[h!]
\centering
\begin{center}
\begin{tabular}{ccccccc}
\hline
\multicolumn{1}{l}{}                  & \multicolumn{2}{c}{FACS}           & \multicolumn{2}{c}{Visemes}               & \multicolumn{2}{c}{Overall} \\ \hline
\multicolumn{1}{l|}{Evaluation-Metric\textbackslash Method}  & Moser et al. & \multicolumn{1}{c|}{Ours}  & Moser et al. & \multicolumn{1}{c|}{Ours}  & Moser et al.     & Ours     \\ \hline  \hline
\multicolumn{1}{l|}{Quantitative (MSE) $\downarrow$} & 40.43        & \multicolumn{1}{c|}{29.95} & 42.73        & \multicolumn{1}{c|}{22.59} & 40.92            & 28.37    \\
    \multicolumn{1}{l|}{Qualitative (MOS) $\uparrow$} & 2.43         & \multicolumn{1}{c|}{4.28}  & 2.38         & \multicolumn{1}{c|}{3.81}  & 2.41             & 4.05   \\
    \hline
\end{tabular}
\label{comp_tab}
\end{center}
\caption{Quantitative and qualitative performance of ours and Moser et al. approaches. The quantitative evaluation represents the MSE of landmarks similarity on 39 videos. The qualitative evaluation represents the subjective results (MOS) of a survey of 80 participants. \\MSE, Mean square error - a quantitative measure; MOS, Mean Opinion Score - a qualitative measure; Downarrow symbol indicates lower is better; Uparrow symbol indicates higher is better.}
\end{table*}

For quantitative evaluation, we introduce the landmarks similarity metric. Herein, the source actor frame and the corresponding rendered 3D character image are compared based on facial landmarks. First, both images are cropped around their face. Next, facial landmarks are extracted from both crops and are processed by landmarks alignment procedure to the same template using an sRT transformation. A mean square error (MSE) is calculated between aligned landmarks, representing the distance between the source actor’s facial expression and the translated expression over the 3D character. Table~\ref{comp_tab} shows that our method outperforms \citep{moser2021semi} method in both qualitative and quantitative metrics.

\subsubsection{Ablation Study}
In this section, we provide ablation experiments substantiating the need for the different grouping layers of the Landmarks to blendshape weights network. The No-Grouping network replaces all layers with a simple MLP model that directly regresses the blendshape weights from the facial landmarks. The Conv-Grouping model only uses our grouping method for the 1D convolution layers (i.e., each facial region has its convolution weights). The convolution features are propagated to the blendshape weights directly via an MLP network eliminating the last grouping layers. The Full-Grouping model contains all grouping layer components. 
In Table \ref{ablation_tab}, all these effects are reported by calculating the MSE between the facial landmarks of the actors to the 3D characters on the test dataset.
One can observe an improvement in accuracy at the introduction of every proposed component with a final reduction of 10\% in the Error over the baseline.

\begin{table*}[h!]
\centering
\begin{center}
\begin{tabular}{l|c|c|c|c}
\hline
\multicolumn{1}{l}{}                  & \multicolumn{1}{c}{}           & \multicolumn{2}{c}{Overall Results}               & \multicolumn{1}{c}{} \\
\hline
Evaluation-Metric\textbackslash Method & Moser et al. & Ours No-Grouping & Ours Conv-Grouping & Ours Full-Grouping \\ \hline \hline
MSE $\downarrow$                       & 40.92        & 31.52       & 28.48         & 28.37    \\ \hline
\end{tabular}
\end{center}
\label{ablation_tab}
\caption{Ablation study for the different grouping
layers of the Landmarks to blendshape weights network. \\MSE, Mean square error - a quantitative measure; Downarrow symbol indicates lower is better.}
\end{table*}

\section{Conclusion}
One of the main challenges of establishing a video retargeting system is acquiring an optimal dataset for the supervised learning approach. Labeling each video frame manually with its corresponding blendshape weights (62 blendshapes in our case) can be laborious, time-consuming, and subjective. A reasonable solution could be training an AI model via a synthetic dataset. Yet, this approach requires an expensive dataset that contains a diverse range of 3D characters rendered in high-quality photo-realistic scenes. In this work, we describe a technique that overcomes these challenges to produce a facial animation model trained with only a single 3D character.

Here, we introduce a full pipeline that benefits from facial landmarks to reduce the domain gap between the synthetic 3D character encountered during training to the real-footage inference actors. This approach eliminates using a large-scale, expensive dataset and enables us to achieve sufficient performance with only one 3D character. 

Through this pipeline, we translate the facial landmarks information into blendshape weights by a unique grouping approach where each spatial region of landmarks is grouped, and the knowledge propagates hierarchically until it reaches the corresponding target shape. We further demonstrate a technique to complete the expressions range by implementing target-shape-specific sub-modules.

We show the effectiveness of our method in various aspects. The proposed pipeline captures the real footage of actors’ facial expressions in both Visemes and FACS frames. We demonstrate the robustness of our method by capturing multiple actors and applying their frames to the pipeline over multiple 3D characters that were excluded from the training procedure. We further compared our results to Moser et al. qualitatively and quantitively, achieving a higher mean opinion score (MOS) by 68\% and a lower mean squared error (MSE) by 30.7\%.

Overall, our work provides a state-of-the-art solution for video retargeting using a single 3D character in a high-level pipeline and low-level deep architecture.

\section*{Acknowledgments}
The authors thank all the human actors and actresses who participated in the work experiments.

\bibliographystyle{plainnat}

\end{document}